\documentclass{aa}
\usepackage{graphicx}
\include{epsf}
\def\lesssim{\mathrel{\hbox{\rlap{\hbox{\lower4pt\hbox{$\sim$}}}\hbox{$<$}}}}

\def\gtrsim{\mathrel{\hbox{\rlap{\hbox{\lower4pt\hbox{$\sim$}}}\hbox{$>$}}}}

\begin{document}

%\thesaurus{07.19.1; 08.03.4; 08.16.5; 09.13.2}

          \title{Warm Molecular Layers in Protoplanetary Disks}

          \subtitle{}

          \author{Y. Aikawa \inst{1}
                 \and
                  G.J. van Zadelhoff \inst{2}
                 \and
                  E.F. van Dishoeck \inst{2}
                 \and
                 E. Herbst\inst{3}
                 }

          \offprints{Yuri Aikawa}

          \institute{Department of Earth and Planetary Sciences, Kobe 
University,
                     Kobe 657-8501, Japan\\
                     email: aikawa@kobe-u.ac.jp
                \and
                     Leiden Observatory, P.O. Box 9513, 2300 RA Leiden,
                     The Netherlands
                \and
                     Departments of Physics and Astronomy, The Ohio State
                     University,\\ Columbus, OH 43210, USA\\
                    }

          \date{Received; accepted}
\titlerunning{Warm molecular layers in disks}

\abstract{We have investigated molecular distributions in
protoplanetary disks, adopting a disk model with a temperature
gradient in the vertical direction.  The model produces sufficiently
high abundances of gaseous CO and HCO$^+$ to account for line
observations of T Tauri stars using a sticking probability of unity
and without assuming any non-thermal desorption.  In regions of
radius $R\gtrsim 10$ AU, with which we are concerned, the temperature
increases with increasing height from the midplane.  In a warm
intermediate layer, there are significant amounts of gaseous molecules
owing to thermal desorption and efficient shielding of ultraviolet
radiation by the flared disk.  The column densities of HCN, CN, CS, H$_2$CO,
HNC and HCO$^+$ obtained from our model are in good agreement with the
observations of DM Tau, but are smaller than those of LkCa15.
Molecular line profiles from our disk models are calculated using a
2-dimensional non-local-thermal-equilibrium (NLTE) molecular-line
radiative transfer code for a direct comparison with observations.
Deuterated species are included in our chemical model.  The molecular
D/H ratios in the model are in reasonable agreement with those
observed in protoplanetary disks. {  }}

\maketitle

\section{Introduction}
It is well established from millimeter and infrared observations that
the birth of solar-mass stars is accompanied by the formation of a
circumstellar disk (Beckwith \& Sargent 1996, Natta et al.\ 2000).
These disks are important both as reservoirs of material to
be accreted onto growing stars and as sites of planetary
formation. Because the gas and dust in the disk are the basic
components from which future solar systems are built, studies of their
chemistry are essential to investigate the link between interstellar
and planetary matter. Moreover, the chemical abundances and molecular
excitation depend on physical parameters in the disks such as
temperature and density, and on processes such as radial and vertical
mixing. Thus, studies of the chemistry in protoplanetary disks can
help to constrain their physical structure.

Although CO millimeter lines are routinely used to trace the gas and
Keplerian velocity field in disks around classical T Tauri stars
(e.g., Kawabe et al.\ 1993, Koerner et al.\ 1993, Dutrey et al.\ 1994,
1996, Koerner \& Sargent 1995, Saito et al.\ 1995, Guilloteau \&
Dutrey 1998, Thi et al.\ 2001), detections of other molecules are
still rare.  Dutrey et al.\ (1997) and Kastner et al.\ (1997) were the
first to report observations of molecules such as HCO$^+$, HCN, CN,
HNC, H$_2$CO, and C$_2$H in the disks around GG Tau, DM Tau and TW
Hya.  Dutrey et al.\ found that the abundances of
these species relative to hydrogen in the DM Tau and GG Tau disks are
lower than those in molecular clouds by factors of 5-100 (Dutrey et
al. 1994, Dutrey et al. 1997, Guilloteau et al. 1999).
The abundance ratio of CN/HCN, on the other hand, is
significantly higher than in molecular clouds in all three disks.
More recently, Qi (2000), van Zadelhoff et al.\ (2001) and Thi
et al.\ (in preparation) have reported observations of molecules other
than CO in the disks around the T Tauri stars LkCa15 and TW Hya, and
in those around the Herbig Ae stars MWC 480 and HD 163296, confirming
the trends of high CN/HCN ratio and molecular depletion; the disk
masses derived from CO (and its isotopes) are significantly smaller
than those estimated from the dust continuum assuming a gas to dust ratio
of 100.

Early models of the chemistry in disks considered mostly the
one-dimensional radial structure in the cold midplane (e.g.
Aikawa et al.\ 1997, Willacy et al.\ 1998). Subsequently, it
has been recognized that the vertical stratification of the molecules
is equally relevant. Aikawa \& Herbst (1999a) investigated the
two-dimensional chemical structure within the so-called
Kyoto disk model, representative of the minimum mass solar
nebula. This model has low midplane temperatures and is isothermal
in the vertical direction. The molecular abundances were found to vary
significantly with height $Z$ from the midplane. At large
radii ($R>$100 AU), the temperature is so low that most species,
except H$_2$ and He, are frozen out onto the grains. This depletion is
most effective in the midplane region ($Z\approx 0$) because of the
higher density and hence shorter timescales for molecules to collide
with and stick to the grains.
In regions above and below the midplane, significant amounts of
molecules can remain in the gas phase for longer periods because of
the lower densities, and because of non-thermal desorption by cosmic
rays and/or radiation (e.g., X-rays) from the interstellar radiation
field and the central star.  Aikawa \& Herbst (1999a) suggested that
the observed molecular line emission comes mostly from this region.
There is a height distinction between stable molecules and radicals,
however.  In the surface region of a disk, radicals such
as CN are very abundant because of photodissociation via ultraviolet
radiation, whereas the abundances of the stable molecules such as HCN
peak closer to the midplane.  The molecular column densities obtained
by integrating over height at each radius compare well with those
derived from observations of DM Tau (Dutrey et al.  1997), although
detailed comparison through radiative transfer (i.e.  calculation of
molecular line intensities from model disks) is still lacking. The
freeze-out of molecules in the midplane explains the low average
abundances of heavy-element-containing species relative to hydrogen,
whereas the high abundance ratio of CN to HCN is caused by
photodissociation in the surface layers.

To obtain this good agreement with observations, Aikawa \& Herbst
(1999a) were forced to use the simplifying assumption that
the probability $S$ for sticking upon collision of a molecule
with a grain is significantly smaller than unity.  If the
sticking probability is unity and if only thermal desorption is
considered, the molecular column densities obtained in the Kyoto model
are much smaller than observed, which suggests that either there is
some efficient non-thermal desorption mechanism, or that the disk
temperature is higher than assumed in the Kyoto model.  Aikawa \&
Herbst (1999a) adopted an artificially low sticking probability
$S=0.03$ in order to reproduce the observed CO spectra, without
specifying the non-thermal desorption process or modifying the
temperature distribution in the Kyoto model.  Since adsorption is such
a dominant process in the disk, the cause of the apparently low sticking
probability should be considered more seriously.

In order to explain the strong mid-infrared emission from disks,
several groups suggested the possibility of higher dust temperatures
than assumed in the Kyoto model due to
efficient reception of stellar radiation by flared disks
(e.g., Kenyon
\& Hartmann 1987, Chiang \& Goldreich 1997, D'Alessio et al.\ 1998,
1999).  In the two-layer Chiang \& Goldreich (1997) model (C-G model
hereafter), the upper layer (the so called ``super-heated'' layer) is
directly heated by  stellar radiation from the central star to
temperatures $T\gtrsim 50$ K at radii of $\sim 100$ AU. Also, recent
observations of high-frequency lines (e.g.\ CO $J=6-5$) support the
possibility that the disk temperature is higher than assumed in the
Kyoto model (Thi et al.  2001, van Zadelhoff et al.\ 2001).  Willacy
\& Langer (2000) investigated the molecular distributions in the C-G model
to see if the super-heated layer can maintain enough gaseous organic
molecules to account for the observations.  It was found, however,
that molecules in this layer are destroyed by the harsh ultraviolet
radiation from the star, whereas they are frozen out onto the grains
in the cold lower layer.  These authors therefore had to adopt a very
high photodesorption rate in the lower layer to keep the molecules off
the grains.

In this paper we report an investigation of molecular distributions in
another disk model with a vertical temperature gradient: the model of
D'Alessio et al.\ (1998, 1999).  These scientists obtained the
temperature and density distribution in steady accretion disks around
T Tauri stars by solving the equations for local 1-D energy transfer
(including radiation, convection and turbulence) and hydrostatic
equilibrium in the vertical direction.  Whereas in the C-G model, the
disk is divided into two discrete layers -- super-heated and interior
-- the model of D'Alessio et al.\ gives continuous distributions of
temperature and density.  The differences in the temperature and
density distributions between these two models have a significant
effect on the gaseous molecular abundances in the disk.  Since a
vertical distribution of density in the super-heated layer is not
given explicitly in the C-G model, Willacy \& Langer (2000) assumed a
Gaussian density distribution with a scale height determined by the
mid-plane temperature.  In the model of D'Alessio et al., the gas is
more extended than in a Gaussian distribution owing to the high
temperature in the surface region.  The higher densities
(and thus the higher column densities)
at large $Z$ shield the lower layers from stellar ultraviolet
radiation. In addition, the temperature variation in the vertical
direction is more gradual in the model of D'Alessio et al.\ than the
step function assumed in the C-G model. Therefore, compared with the
C-G model, the model of D'Alessio et al.\ contains more gas in a warm
and shielded layer, in which high abundances of gaseous molecules are
expected.

The rest of the paper is organized as follows.  In \S 2 we describe
the adopted model for protoplanetary disks and the chemical reaction
network.  Numerical results on the distributions of molecular
abundances and column densities are discussed in \S 3.  In \S 4,
molecular column densities and line intensities in the D'Alessio et al.\
model are compared with observations.
Our conclusions and a discussion are given in \S 5.

\section{Model}

\begin{figure*}
\resizebox{14cm}{!}{\includegraphics{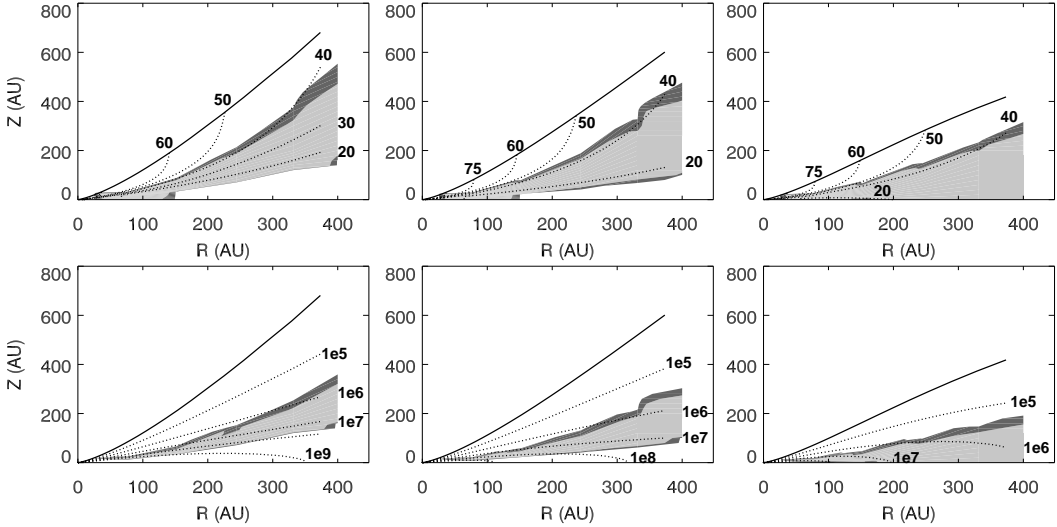}}
%\hfill
%\parbox[b]{55mm}{
\caption{The distributions of temperature $T$ [K] (top row) and density
$n$(H$_2$) [cm$^{-3}$] (bottom row) in the D'Alessio et al.\ model with
a viscosity parameter $\alpha = 0.01$ are plotted as dotted lines.
The top thick line in each panel shows the
upper boundary of the model defined by $P=10^{-10}$ dyne cm$^{-2}$, which
corresponds to a gas density of $n$(H$_2$)$=1.4\; 10^4$ cm$^{-3}$ with
$T=50$ K.
  From left to right the mass accretion rate
is $10^{-7}, 10^{-8}$, and $10^{-9}$ M$_{\odot}$ yr$^{-1}$, respectively.
In the top row, the grey scale
shows the region in which the CO abundance relative to hydrogen is $10^{-6}-
10^{-5}$ (dark grey) and $10^{-5}-10^{-4}$ (light grey). In the bottom row,
the fractional HCO$^+$ abundance is $10^{-11}-10^{-10}$ (dark grey)
and $10^{-10}-10^{-9}$ (light grey).}
\label{fig:2Dmodels}
%}
\end{figure*}

The disk model by D'Alessio et al.\ (1998, 1999) has been adopted in
our work.  In this model, the two-dimensional structure ($R, Z$) of the disk
is obtained by solving for hydrostatic equilibrium and energy transport in the
$Z$-direction.
Various heating sources are considered, such as viscous dissipation
of accretion energy, cosmic rays, and stellar radiation. Among them,
stellar radiation is dominant at $R>2$ AU, assuming typical parameters of T
Tauri stars: $T_{\ast}=4000$~K, $M_{\ast}=0.5$ M$_{\odot}$, and
$R_{\ast}=2$ R$_{\odot}$.  We adopt a disk with an accretion rate
$\dot{M}=10^{-8}$ M$_{\odot}$ yr$^{-1}$ and viscosity parameter
$\alpha=0.01$ as the fiducial, or standard, model,
but also consider cases with the same $\alpha$ and differing accretion rates
$\dot{M}=10^{-7}$ M$_{\odot}$ yr$^{-1}$
and $\dot{M}=10^{-9}$ M$_{\odot}$ yr$^{-1}$
(D'Alessio et al. 1999).
The distributions of density and temperature in
these three models are shown in Fig. \ref{fig:2Dmodels}.  The
surface density in the fiducial model is similar to that in the
minimum-mass disk, which was adopted by Aikawa \& Herbst (1999a), and is
approximately an order of magnitude larger (smaller) in the model with the
larger (smaller) $\dot{M}$. In the fiducial disk, the masses inside 100 AU
and 373 AU are about 0.017 M$_{\odot}$ and 0.06 M$_{\odot}$, respectively.

We have selected a few radial points from the model of D'Alessio et
al.\ (viz.  $R=$ 26, 49, 105, 198, 290, and 373 AU), divided the
disk at each radius into several (30-40) layers, or slabs, depending
on height $Z$, and calculated the molecular abundances in each layer
as functions of time (Aikawa \& Herbst 1999a).  We have not included
any hydrodynamic motions in the disk, such as accretion or turbulence.
The main goal of this paper is to investigate the effect of a vertical
temperature gradient on molecular abundances and line intensities.
Although the model of D'Alessio et al.\ is an accretion disk model,
the temperature distribution is determined by the irradiation from the
central star and radiation transfer in the vertical direction, while
the contribution of the accretion energy as a heat source is
negligible in the region we are concerned with.  In addition, we find
that the chemical time scale is shorter than the accretion timescale,
which is $\sim 10^6$ yr, in a large fraction of the molecular
layers (section 3.2), so that molecular distributions obtained in this paper
%do not depend sensitively on the parameters.
can be a reasonable approximation of reality.

The chemical model and chemical reaction equations adopted in this paper
are almost the same as those described in Aikawa \& Herbst (2001).  We use the
``new standard model'' for the gas-phase chemistry (Terzieva \& Herbst 1998,
Osamura et al.\ 1999), extended to include deuterium chemistry (Aikawa \&
Herbst 1999b).  The ionization rate by cosmic rays is assumed to be
the ``standard'' value in molecular clouds, $\zeta =1.3\; 10^{-17}$
s$^{-1}$, because the attenuation length for cosmic-ray ionization
is much larger than the column densities in the outer regions of the
disks, with which we are concerned (Umebayashi \& Nakano 1981).
Photoprocesses induced by
ultraviolet radiation from the interstellar radiation field and from
the central star are included.  The ultraviolet flux from the central
star varies with time and object, and can reach a value $10^4$
times higher than the interstellar flux at $R=100$ AU (Herbig \& Goodrich
1986, Imhoff \& Appenzeller 1987, Montmerle et al.  1993).
This maximum value is adopted in this paper, as in Aikawa \& Herbst (1999a).
We assume that the ultraviolet radiation from the central star is not energetic
enough to dissociate CO and H$_2$.  Self- and mutual shielding of
H$_2$ and CO from interstellar UV is considered as in Aikawa \& Herbst
(1999a). Chemical processes induced by X-rays from the central
star are not
included in this paper.

Regarding gas-grain interactions, the surface formation of H$_2$, the
surface recombination of ions and electrons, and the accretion and
thermal desorption of ice mantles are included, but all other
grain-surface reactions are not included.  The sticking probability
for accretion is assumed to be 1.0, unless stated otherwise.
We have not considered any
modifications to the grain-surface rate equation for H$_2$ formation
(cf.\ Caselli et al.\ 1998), because, given our initial conditions,
the rate of molecular hydrogen formation is not important to the
model.  The total number of species and reactions included in our
network are 773 and 10446, respectively.

The adopted elemental abundances are the so-called ``low-metal''
values (e.g., Lee et al.\ 1998, Aikawa et al.\ 1999).  The initial
molecular abundances are obtained from a model of the precursor
cloud with physical conditions $n_{\rm H}=2\; 10^4$
cm$^{-3}$ and $T=10$ K at $3\; 10^5$ yr, at which time observed
abundances in pre-stellar cores such as TMC-1 are reasonably well
reproduced (Terzieva \& Herbst 1998).

\section{Results}
\subsection{Vertical Distribution}

\begin{figure*}
\resizebox{14cm}{!}{\includegraphics{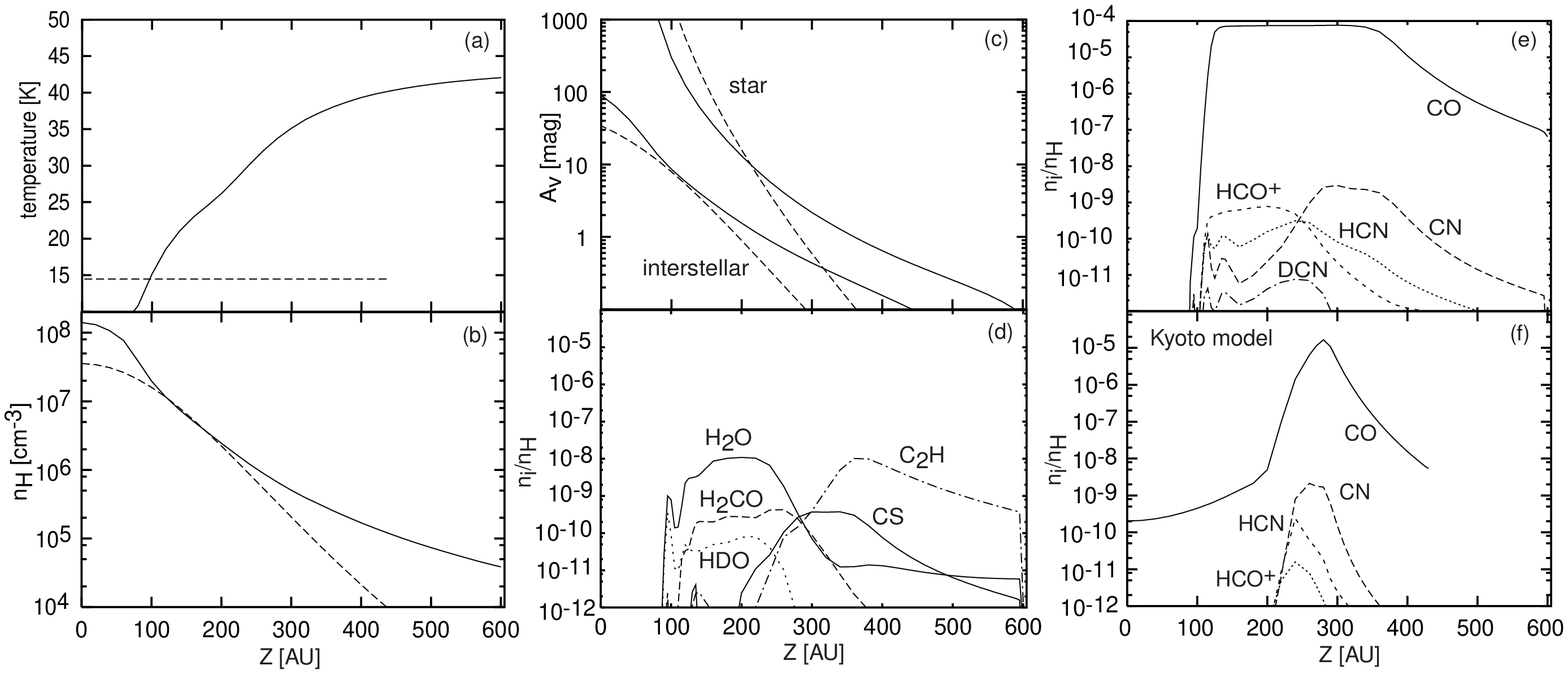}}
%\hfill
%\parbox[b]{55mm}{
\caption{Vertical distributions at $R=373$ AU of $(a)$ temperature,
$(b)$ density ($n_{\rm H}\equiv 2 n({\rm H}_2)+n({\rm H})$),
$(c)$ attenuation of the interstellar radiation ($A_{\rm v}^{\rm IS}$)
and stellar radiation ($A_{\rm v}^{\rm star}$), $(d-e)$ molecular abundances
in the D'Alessio et al.\ model, and $(f)$ molecular abundances in the Kyoto
model with $S=0.03$.
In panels ($a-c$), the physical parameters of the D'Alessio et al.\
and Kyoto models are shown via {\em solid} and {\em dashed} lines,
respectively.
In panels ($d-f$), the disk age is assumed to be $t=1.0\; 10^6$ yr.}
\label{fig:vertical_rel}
%}
\end{figure*}

Figure 2 contains assorted vertical distributions at the outermost radius
$R=373$ AU in our fiducial disk model.
The solid lines in Fig. \ref{fig:vertical_rel} ($a-c$) show the physical
parameters: temperature, density, and A$_{\rm v}$.
The attenuation of interstellar radiation (A$_{\rm v}^{\rm IS}$) is obtained
from the equation
\begin{equation}
A_{\rm v}=\frac{N_{\rm H}}{1.59\ 10^{21} {\rm cm}^{-2}} \ \ {\rm mag},
\end{equation}
where $N_{\rm H}$ is the vertical column density of hydrogen
nuclei from the disk
surface to each point in the disk. The attenuation of stellar radiation
(A$_{\rm v}^{\rm star}$) is obtained via the same equation,
but with $N_{\rm H}$ replaced by the column density from the central star.
Fig. \ref{fig:vertical_rel} ($d-e$) shows
assorted molecular abundances at a disk age of $t=1.0\; 10^6$ yr,
which is typical for T Tauri stars.  Significant amounts of molecules
exist in the gas phase due to thermal desorption at $Z\gtrsim 100$ AU,
while most molecules are adsorbed onto grains below this height, where
$T\lesssim 20$ K.
As discussed by van Zadelhoff et al.\ (2001), this molecular layer
covers the region in which the lines of the main isotopes of
the observed species become optically thick and thus where most of the
observed emission arises. The results for models with different
$\dot{M}$ are similar
except that the height of the molecular layer is shifted in accordance
with the distribution of $A_{\rm v}^{\rm IS}$, which is the main
determinant of the vertical temperature distribution.

The vertical distribution of the physical parameters (dashed lines)
and molecular abundances in the Kyoto model are also shown in Fig.
\ref{fig:vertical_rel} ($a-c$, and $f$) for comparison.  The mass of
the central star in the latter model is 0.5 M$_{\odot}$, as in the
D'Alessio et al.\ model, so that the density distribution is modified
from that adopted in Aikawa \& Herbst (1999a). The sticking
probability of neutral species onto grain surfaces is assumed to be
0.03, as in Aikawa \& Herbst (1999a). It can be seen that the density
distribution in the D'Alessio et al.\ model is more extended than the
Gaussian profile assumed in the Kyoto model (as well as in Willacy \&
Langer 2000), causing more efficient ultraviolet
shielding of the warm layers just below the surface.
Although we have not calculated the molecular distributions in the C-G
model, the width of its warm molecular layer at $R=373$ AU can be estimated.
% if desorption is only thermal
In the C-G model, the Gaussian density distribution is similar to that in the
Kyoto model, but the disk surface with $A_{\rm v}^{\rm star}
\lesssim 1$ mag is ``super-heated'' by stellar radiation. At $R=373$ AU,
the boundary between the super-heated layer and the interior region
   is located
at $Z\sim 280$ AU, a value estimated from the $Z-$A$_{\rm v}^{\rm star}$
relation in the Kyoto model (Fig. \ref{fig:vertical_rel} $c$).
Since the temperature in the interior region is lower than 20 K at this radius
in the C-G model, warm gaseous molecules exist only at heights larger than
280 AU. The upper boundary of the molecular
layer, which is determined by photoprocesses, is estimated to be $Z\sim 350$
AU, again from the Kyoto model (Fig. \ref{fig:vertical_rel} $f$).
Therefore in the C-G model, the warm molecular layer, if any,
is much narrower than in the D'Alessio et al.\ model.
This estimate is consistent with the conclusion of Willacy \& Langer (2000)
that they need non-thermal desorption in the cold, more shielded
layer to account for the observed molecular
abundances within the C-G model.

\subsection{Radial Distribution of Column Densities}

\begin{figure*}
\resizebox{14cm}{!}{\includegraphics{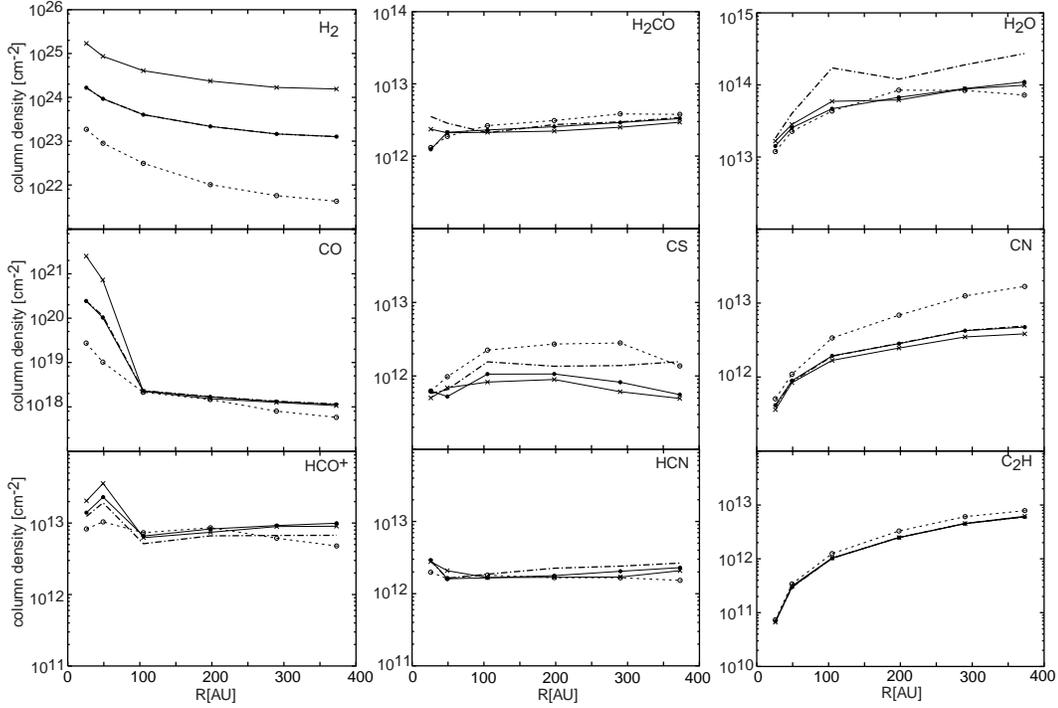}}
%\hfill
%\parbox[b]{55mm}{
\caption{The column densities of assorted molecules as functions of radius in
models with mass accretion rate  $1.0\; 10^{-7}$ ({\em solid lines with
crosses}), $1.0\; 10^{-8}$ ({\em solid lines with closed circles}), and
$1.0\; 10^{-9}$ ({\em dashed lines with open circles})
M$_{\odot}$ yr$^{-1}$.
The viscosity parameter is fixed at $\alpha=0.01$, while the disk age
is $t=1.0\;10^6$ yr. The column densities at $t=1.0\;10^5$ yr in the model\
with mass accretion rate $1.0\; 10^{-8}$ M$_{\odot}$ yr$^{-1}$
are shown with
{\em thick dot-dashed lines}.}
\label{fig:column}
%}
\end{figure*}

Column densities are obtained by integrating the molecular abundances
in the vertical direction.  The column densities of assorted species
as functions of disk radius are shown in Fig.  \ref{fig:column} for
three accretion disk models at $t=1.0\;10^6$ yr with different mass
accretion rates.  Stable neutrals such as H$_2$CO and HCN show little
dependence on radius, because these molecules are abundant only in
regions with certain physical conditions and the mass contained in the
layer with these physical conditions does not vary much with radius.
For example, the HCN abundance is high ($n$(HCN)/$n_{\rm H} \sim
10^{-10}-10^{-9}$) only when $n_{\rm H}\lesssim 3 \; 10^7$
cm$^{-3}$, $T\gtrsim$ 20 K, and $A_{\rm v}^{\rm IS}\gtrsim 0.2$ mag
(see Fig.  \ref{fig:vertical_rel}).  The critical temperature of $\sim
20$ K is not the sublimation temperature of HCN, but that of CO, which
is the dominant form of carbon in the gas phase. For HCN, attenuation of
interstellar radiation is more important than that of stellar
radiation because the interstellar radiation penetrates deeper into
the disk due to the effect of geometry. Radicals such as CN and
C$_2$H increase in column density with radius because of the lower
density and lower flux of the destructive stellar UV in the outer regions.
Radical column densities are more sensitive than HCN to stellar UV, because
their abundances peak at greater heights.
%{\it EvD: I suggest to leave out the next sentence: Radical
%species can survive in a harsher UV field than can other neutral
%species such as HCN, but direct stellar UV at the very surface region
%is too destructive even for them to maintain a high abundance.
Carbon monoxide is
abundant ($n$(CO)/$n_{\rm H} \sim 10^{-4}$) in regions with $T\gtrsim
20$ K and $A_{\rm v}^{\rm IS} \gtrsim 0.1$ mag.  The column densities
of CO and HCO$^+$ abruptly change at $R\sim 100$ AU, inside of which
the temperature in the midplane is higher than 20 K. These
characteristics of the radial distribution are similar to those in the
Kyoto model (Aikawa et al.\ 1996, Aikawa \& Herbst 1999a).

The amount of gas existing under physical conditions conducive for
large abundances of molecules does not vary significantly among
the three disk models with different accretion rates (and thus with
different disk mass), either.  Therefore, most (but not all) molecular
column densities vary only by a very small factor among the three disk
models, even though the total (H$_2$) column density varies by two
orders of magnitude.  In the model with a mass accretion rate of
$1.0\; 10^{-9}$ M$_{\odot}$ yr$^{-1}$, the region with density $n_{\rm
H}\sim 10^5-10^6$ cm$^{-3}$, at which CN is abundant, is more
shielded from stellar UV (Fig.
\ref{fig:2Dmodels}), and thus the CN column density is higher than in
the other two models.

Molecular column densities in our fiducial disk model at an
earlier time of $t=1.0\;10^5$ yr are shown with thick dot-dashed
lines in Fig.  \ref{fig:column}.
The variation in column density for
most molecules during $10^5-10^6$ yr is less than a factor of 2; two
exceptions are CS and H$_2$O. In regions with $A_{\rm v}^{\rm IS}$
smaller than a few mag, which covers a large fraction of the molecular
layer, the chemical timescales are short ($\lesssim 10^5$ yr) because
of photoprocesses.  In the more shielded portion of the molecular
layer at smaller $Z$,
S-bearing gaseous molecules decrease in abundance after $10^5$ yr,
since most sulfur is adsorbed onto grains in the form of CS, SO and OCS.
Similarly, H$_2$O gas decreases at $\sim 10^6$ yr, because most oxygen
which is not in CO is adsorbed as H$_2$O ice.  On the other hand,
abundances of other C-bearing molecules reach pseudo-steady-state
values in a relatively short timescale ($\lesssim 10^4$ yr), and do
not show significant time variation during $10^5-10^6$ yr in the more
shielded region because CO gas, their chemical precursor, remains the
dominant component of carbon for more than $ 1\; 10^6$ yr.  The
pseudo-steady-state gas-phase chemistry of non-volatile
carbon-containing species is balanced for a considerable period by
formation reactions starting from CO and depletion onto the dust
particles.

Values of the sticking probability $S$ are estimated to lie in
the range $0.1\lesssim S\lesssim 1.0$ (Williams 1993, and references
therein).  In order to check the dependence of the molecular column
densities on $S$, we have performed calculations with $S=0.1$ in
addition to our fiducial value of unity.  Column densities of radicals
such as CN and C$_2$H do not depend on $S$, because they are abundant
in the surface layer, in which adsorption is not the dominant process.
Among the more stable species, some show significant dependence on
$S$; the column densities of CO$_2$ and OCS are larger by an order of
magnitude in the model with the lower sticking probability at $R=373$ AU
and $t=1\; 10^6$ yr.  But the effect of $S$ is smaller for many other
species; at $R=373$ AU and $t=1\; 10^6$: for example, the column
densities of CS, HCN, and H$_2$CO are larger only by factors of 3.2,
1.7, and 1.3, respectively, in the case of lower $S$.  There are two
reasons for this small dependence on $S$.  First, adsorption
is not always the dominant process in
the molecular layer, depending on species and height from the
midplane.  Second, the higher abundance of gaseous O$_2$ in the case
of lower $S$ reduces the abundance of the carbon atom, and thus
reduces the formation rate of organic molecules, which counteracts the
lower adsorption rate.

\subsection{D/H Ratios in Molecules}

\begin{figure*}
\resizebox{15cm}{!}{\includegraphics{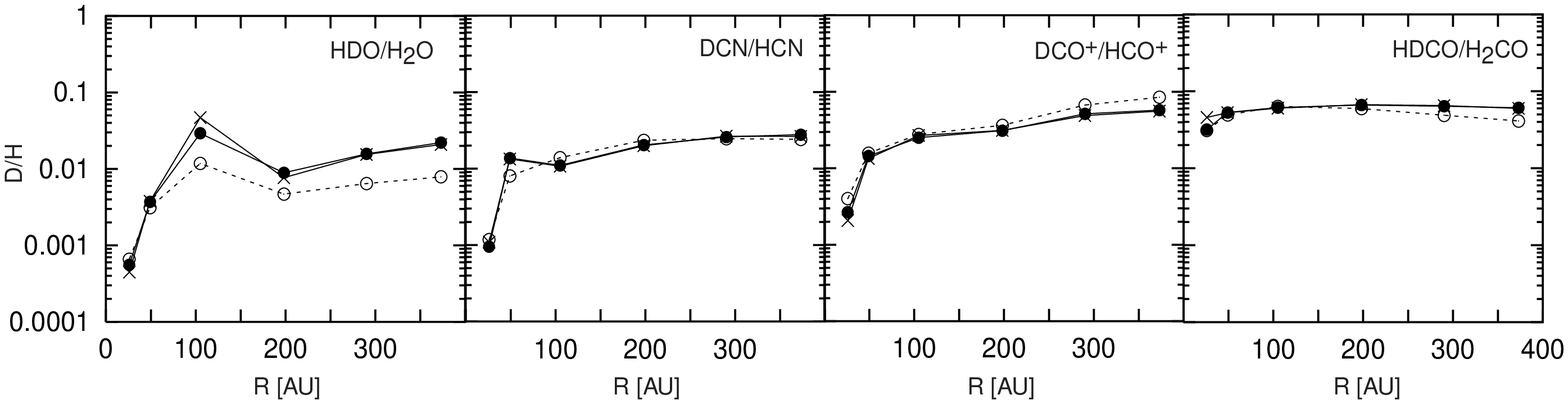}}
%\hfill
%\parbox[b]{55mm}{
\caption{Column density ratios of deuterated species to normal species
as functions of radius in the three accretion disk models at an age of
$t=1.0\;10^6$ yr. The models are represented by the same lines as in
the previous figure.}
\label{fig:DH}
%}
\end{figure*}

In the disks around LkCa15 and TW Hya (Qi 2000, Thi et al. in preparation),
the deuterated species DCN and DCO$^+$ have been detected, and
the ratios of deuterated to normal species (DCN/HCN and DCO$^+$/HCO$^+$)
are estimated to be $\sim 0.01$. The deuterated species HDO has
been detected in LkCa15, but there is no estimated ratio of HDO to normal
water. Although the estimated ratios might include
uncertainties as discussed in \S 4, they would not alter the important
conclusion that the molecular D/H ratios are higher than the cosmic elemental
abundance ratio (D/H$\approx 1.5\; 10^{-5}$) by orders of magnitude.
It is interesting to see if our model
reproduces the high D/H ratios and if these ratios can be used as
probes of the chemical processes and physical conditions in
disks. Fig. \ref{fig:DH} shows some column density ratios of
deuterated species to normal species in our models as functions of
radius. The ratios are similar to the observed values of 0.01
for both HCN and HCO$^+$.
The mechanism of deuterium enrichment in disks
is similar to that in molecular clouds; due to energy differences
between deuterated and normal species, molecules such as H$_3^+$ have
a high D/H ratio, which propagates to other species through chemical
reactions (Millar, Bennet \& Herbst 1989, Aikawa \& Herbst 1999b,
Roberts \& Millar 2000). The
D/H ratios decrease as the temperature rises, because the energy
differences are less significant at higher temperatures.  Thus, the
D/H ratio decreases inwards (Fig. \ref{fig:DH}), because the
temperature in the gaseous molecular layer is higher at inner radii.
Because the temperature in the molecular layer does not depend much on the
disk mass (see Fig. \ref{fig:2Dmodels}), the differences between
models with different $\dot{M}$ are small.

The ratio of HDO/H$_2$O shows a more complicated behavior than
described above.  It has a local peak at $R\sim100$ AU, and the peak
is higher in the more massive disk model.
At $R\sim 100$ AU, the midplane is at 14-16 K for the three disk models,
at which temperature CO is almost completely frozen
onto grains but O can marginally be kept in the gas phase to produce
water vapor.  The D/H ratio in the vapor is enhanced by the CO
depletion in the region, since CO is one of the main destruction
channels of H$_2$D$^+$ (Brown et al.\ 1989). The layer at this critical
temperature (14 K$\lesssim T \lesssim 16$ K) is thicker in the more massive
disk.  At smaller radii, the midplane temperature is higher, which
lowers the D/H ratio. At larger radii for disks with $\dot{M}=10^{-7}$
and $10^{-8}$ M$_{\odot}$ yr$^{-1}$, the midplane temperature is so low
that O cannot remain in the gas phase to produce water vapor. The
abundance of HDO has a sharp peak in a thin layer of 14 K
$\lesssim T \lesssim 16$ K
offset from the midplane (Fig. \ref{fig:vertical_rel} $d$).  In the outer
disk of the $\dot{M}=10^{-9}$ M$_{\odot}$ yr$^{-1}$ model, the
temperature is slightly higher than 20 K even in the midplane (see
Fig. \ref{fig:2Dmodels}), which causes a lower HDO/H$_2$O ratio than
for the other two models.

\begin{table}
\begin{tiny}
       \caption[]{Calculated ($t=1.0\;10^6$ yr) molecular column densities
(cm$^{-2}$) at $R=373$ AU compared with observations}
          \label{tab:column}
      $$
          \begin{array}{l c c c c c c}
           \hline
           \noalign{\smallskip}
            {\rm Species} & \multicolumn{3}{c}{\rm \dot{M} \;
[M_{\odot} yr^{-1}]} &{\rm DM Tau}^{\mathrm{a}} &
\multicolumn{2}{c}{\rm LkCa15}\\
            & {\rm 10^{-7}} & {\rm 10^{-8}} & {\rm 10^{-9}} & &
{\rm interferometer}^{\mathrm{b}} & {\rm single\; dish}^{\mathrm{c}} \\
           \noalign{\smallskip}
           \hline
           \noalign{\smallskip}
           {\rm H_2}    & 1.5(24)^{\mathrm{d}}   & 1.3(23)  & 4.3(21)   &
           3.8(21) & & \\
           {\rm CO}     & 1.1(18)   & 1.1(18)  & 5.8(17)   &
5.7(16)^{\mathrm{e}} & 1.6(18)^{\mathrm{e}}   & 9.0(17)^{\mathrm{f}} \\
           {\rm HCN}    & 2.1(12)   & 2.3(12)  & 1.5(12)   &  2.1(12) &
0.02-1.2(15)^{\mathrm{g}} & 7.8(13) \\
           {\rm CN}     & 3.8(12)   & 4.7(12)  & 1.7(13)   &  9.5-12(12)&
9.1-25(13) & 6.3(14)\\
           {\rm CS}     & 4.9(11)   & 5.6(11)  & 1.4(12)   &  6.5-13(11)
& 1.9-2.1(13) & 2.2(14)\\
           {\rm H_2CO}  & 2.9(12)   & 3.3(12)  & 3.8(12)   &  7.6-19(11) & &
3.0-22(13)\\
           {\rm HCO^+}  & 9.0(12)   & 9.9(12)  & 4.8(12)   &  4.6-28(11) &
           1.5(13) & 1.4(13)\\
           {\rm C_2H}   & 6.2(12)   & 6.0(12)  & 7.9(12)   &  4.2(13)
& &\\
           {\rm HNC}    & 2.0(12)   & 2.3(12)  & 1.5(12)   &  9.1(11) & <
5.4(12) & \\
           {\rm OCS}    & 3.1(10)   & 2.8(10)  & 5.0(9)    &          & <
2.9(13) & \\
           {\rm CH_3OH} & 6.4(8)    & 7.1(8)   & 6.6(8)    &          &
7.3-18(14) & < 9.4(14) \\
           {\rm DCN}    & 5.5(10)   & 6.4(10)  & 3.7(10)   &          &
1.0(13) & \\
           {\rm HDO}    & 2.1(12)   & 2.4(12)  & 5.7(11)   &          &
2.3-6.8(14) &  \\
           {\rm N_2H^+} & 1.9(12)   & 1.9(12)  & 6.0(9)    & < 7.6(11)&
< 5.7(12) & < 5.9(13)\\
           \noalign{\smallskip}
           \hline
        \end{array}
      $$
\begin{list}{}{}
\item[$^{\mathrm{a}}$] Derived from single-dish data by Dutrey et al.\ (1997)
(see text).
\item[$^{\mathrm{b}}$] Derived from interferometer data by Qi (2000).
The values in this column do not necessarily refer to 373 AU (see text).
\item[$^{\mathrm{c}}$] Derived from single-dish data by Thi et al.
(in preparation) assuming a disk radius of 100 AU (see text).
\item[$^{\mathrm{d}}$] $a(b)$ means $a\; 10^b$.
\item[$^{\mathrm{e}}$] Estimated from C$^{18}$O assuming
C$^{16}$O/C$^{18}$O$=500$.
\item[$^{\mathrm{f}}$] Estimated from $^{13}$CO assuming
$^{12}$CO/$^{13}$CO$=60$.
\item[$^{\mathrm{g}}$] Lower value is estimated from H$^{12}$CN and higher
value is an upper limit estimated from H$^{13}$CN assuming
H$^{12}$CN/H$^{13}$CN$=60$.
\end{list}
\end{tiny}
       \end{table}

\begin{figure*}
\resizebox{15cm}{!}{\includegraphics{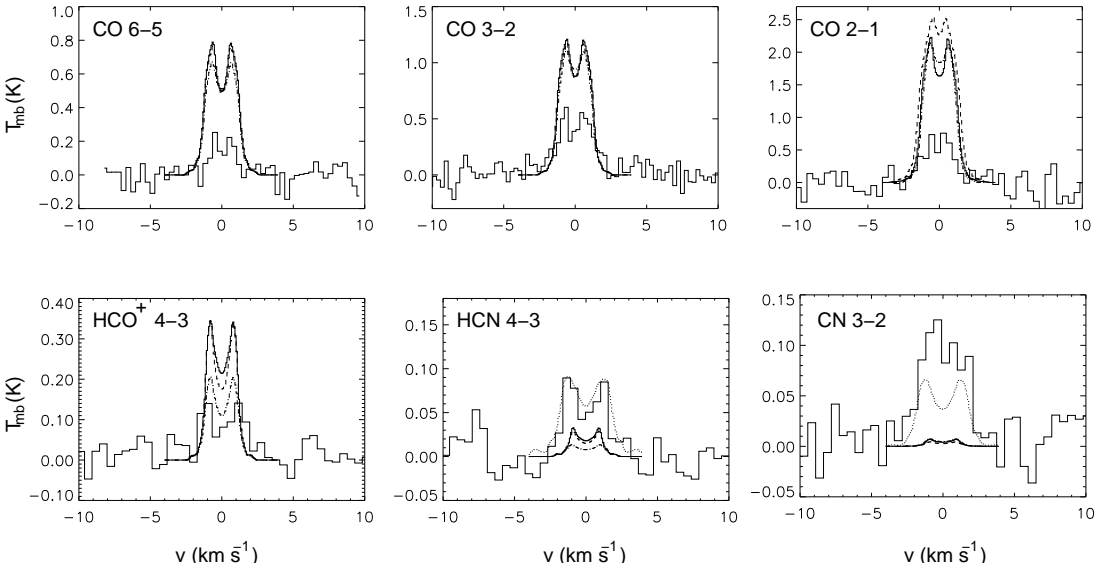}}
%\hfill
%\parbox[b]{55mm}{
\caption{Molecular line profiles from LkCa15 ({\em solid line}),
and the D'Alessio et al.\ model with accretion rates of $1.0\; 10^{-9}$
({\em dot-dashed line}), $1.0\; 10^{-8}$ ({\em thick solid line)}, and
$1.0\; 10^{-7}$ ({\em dashed line}) M$_{\odot}$ yr$^{-1}$.
The dotted line in the HCN profile represents the disk with accretion rate
of $1.0\; 10^{-8}$ M$_{\odot}$ yr$^{-1}$, but HCN abundance
at each position in the disk
is artificially enhanced by a factor of 10 from our calculated values.
Similarly, in the CN profile abundance is enhanced by a factor of
50 for the dotted line. The disk radius, age,
and inclination are 400 AU, $1\; 10^6$ yr and 60 degrees,
respectively.}
\label{fig:line400}
%}
\end{figure*}

\section{Comparison with observation}

\subsection{Column densities}

In the subsequent paragraphs, we discuss comparisons of our calculated
column densities for assorted species at a particular radius with 
both single dish and
interferometric data.  It must be recognized, however, that it is not
really possible to derive reliable columm densities at a particular
radius from unresolved single dish data, making much of the subsequent
discussion less quantitative than would be desirable.

Table \ref{tab:column} compares calculated molecular column densities
obtained with three different accretion rates at a time of
$t=1.0\;10^6$ yr and a radius of $R=373$ AU with those estimated from
the observations of DM Tau (Dutrey et al.  1997) and LkCa15
(Qi 2000, Thi et al. in preparation).
In general, it is difficult to estimate the total H$_2$ column density
in the disk, especially at the outer radius.
Dust observations suffer from uncertainties in
the dust opacity at millimeter wavelengths, which depends on the
grain size distribution. Moreover, the dust continuum is very weak at
the outer radii and difficult to
detect. The H$_2$ column density cannot be directly estimated from
molecular lines, because the molecular abundances relative to hydrogen
are not known.
Therefore Dutrey et al.\ (1997) estimated the H$_{2}$ column density
and averaged molecular abundances for DM Tau in a different way,
paying attention to the critical density for excitation of the
molecular lines and their optical depth.  For the DM Tau disk,
interferometric observations of $^{12}$CO ($J=2-1$) show that the CO
gas extends to $\sim 800$ AU from the central star.  Combining the
different constraints, they obtained an H$_2$ density $n(R,Z)=5\ 10^5
(R/500AU)^{-3}$ exp$[-(Z/H)^2]$ cm$^{-3}$ over this range, in which
$H$ is the scale height of the disk, with $H\approx$ 175 AU at $R=500$ AU.
  From the line intensities obtained in single dish telescopes,
they subsequently derived
molecular abundances with respect to hydrogen assuming that the
abundances are constant over the entire disk. We obtained the
molecular column densities of DM Tau in Table \ref{tab:column} by
vertically integrating this disk model using the average abundances
listed in Table 1 of Dutrey et al. (1997).

Although the DM Tau disk extends to about 800 AU from
the central star, and the outer radius would have a larger contribution to
the line intensities because of the larger surface area, we
list the calculated column density at $R=373$ AU, since the D'Alessio
et al.\ model
extends only out to this radius (the H$_2$ column density at $R=800$ AU
is about 3 times less than that at $R=373$ AU in the DM Tau disk).
The model with an accretion rate $\dot{M}=10^{-9}$ M$_{\odot}$ yr$^{-1}$
reproduces reasonably well the vertical H$_2$ column density of DM Tau,
and agrees with other DM Tau
observations to within a factor of 2, except for CO and C$_2$H. The
former is overestimated, while the latter is underestimated.  It
should be noted that we have assumed that the stellar UV is not hard
enough to dissociate CO, so that our model may overestimate the CO
column density.  Inclusion of CO photodissociation by stellar UV
might also increase C$_2$H and CN abundances, since more
carbon is released from CO in the photodissociation region, in
which C$_2$H and CN are mainly formed (van Zadelhoff et al.
in preparation).

Qi (2000) performed interferometric observations on LkCa15 and
estimated beam averaged molecular column densities based on the
velocity-integrated intensity measured over a much smaller beam than
used to determine the DM Tau abundances.  The vertical column
densities are lower than the listed values by a small factor, which is
less than 2 if the inclination is $\lesssim 60$ degrees. Since
the beam size is about $0.6''-13''$ depending on the frequency of the
line, the estimated values do not necessarily refer to 373 AU.  If the
emission is resolved and if we assume that the molecular column
densities do not vary much with radius (as indicated by our disk
models for $R\gtrsim 100$ AU), we can take the values listed by Qi
(2000) to apply to 373 AU.  The typical beam size of Qi (2000) is
$3''-4''$ ($\sim$300 AU radius at the distance of LkCa15), so the
majority of the results will be just resolved, making it reasonable if
not perfect to compare the observations with our result at $R=373$
AU.

In addition to the interferometric work on LkCa15, Thi et al. 
(in preparation) derived
molecular column densities for this disk from high-frequency
single dish observations of various species. Their beam-averaged values
obtained with their assumption that the disk has a radius of
100 AU are included in Table 1 and differ from the
interferometric column densities by up to an order of magnitude.
Both values are much higher than the column densities found for DM Tau,
%The column densities are much higher than those in DM Tau,
%because the emission is much less extended,
so the agreement between our model and observations is worse for
LkCa15 than for DM Tau.  For CO and HCO$^+$ the difference is 
less than a factor of 2, but the column densities of other species
are about 1--2 orders of magnitude higher than the model results for
all three mass accretion rates.  Methanol in LkCa15 appears to
be a special case; its calculated column density is more than five
orders of magnitude too low, presumably because it is produced by the
hydrogenation of CO on grain surfaces, which is not included in our
model, and then evaporated into the gas.

The use of 100 AU for LkCa15 as a reference point by Thi et
al. is somewhat arbitrary. If the values obtained by Thi et al.\ are
taken to refer to a 373 AU radius disk, their LkCa15 column densities
need to be reduced by a factor (100/373)$^2$, and become closer to
those found for DM Tau. Thi et al.\ and Qi (2000) also present data
for a few other disks (TW Hya, HD 163296, and MWC 480) and determine
column densities and abundances in a consistent way. Indeed, the DM
Tau column densities and abundances are generally lower than the
values for other disks, whereas those for LkCa15 are among the
highest. Thus, these two sources appear to bracket the range of
observed values.

%{\bf Kastner et al. (1997) detected $^{13}$CO, HCN, CN and
%HCO$^+$ towards TW Hya. The relative abundances of the later species
%to $^{13}$CO are similar to those in DM Tau. But we cannot judge if
%our model is consistent with TW Hya, because the column density of
%$^{13}$CO is not estimated.}

\subsection{Line intensities and profiles}

In addition to the comparison with estimated molecular column
densities at a single radius, a more direct comparison via
line intensities from the entire model disk has been made.
Since the most complete single-dish and interferometer data set is available
for LkCa15, we restrict our efforts to this source.
In estimating the
molecular column densities from the observations, Qi (2000) assumed
local thermal equilibrium (LTE) with an excitation temperature of 40 K
throughout the LkCa15 disk, thereby deriving a mean column over the
beam.
In this paper, however, we have shown that a temperature
gradient is important for the characteristics of the gaseous molecular
layer in the disk and that the abundances are strongly varying with
$R$ and $Z$. Also, the excitation of the molecules and the line
emission depend on the density structure; van Zadelhoff et al.\
(2001) have shown that the assumption of LTE is not always valid for high
frequency lines of molecules with
high critical densities.  Hence, it is better to compare directly
simulated line emission with observations.

We calculate here the excitation of the molecules using the 2-dimensional
(2D) NLTE molecular line radiative transfer code of Hogerheijde \&
van der Tak (2000). From the resulting level populations, the line
profiles can be computed, taking into account the inclination of the
source. NLTE molecular line radiative transfer in more than one
dimension has been used in star-formation research only recently
(e.g., Park \& Hong 1995, Juvela 1997). The need for a full treatment
of the radiative transfer in disks follows from the non-locality of
the problem. The level populations depend both on the local parameters
(temperature, density and radiation field) and on the global radiation
field, which in turn depends heavily on the optical depth of the
medium. The main problem is the slow convergence of the level
populations. The adopted code by Hogerheijde \& van der Tak (2000)
uses an accelerated Monte Carlo method to enhance convergence.  A more
elaborate discussion of the methods is given in van Zadelhoff et
al. (2001).

The main input parameters to the radiative transfer
calculations in addition to the molecular data (collision rates,
Einstein A coefficients) are the 2D distributions of temperature,
density and abundance of the molecule of interest at each position ($R,Z$).
The latter is taken directly from the chemical models at a chosen
age. Other important parameters are the turbulent width, taken to be
0.2 km s$^{-1}$ (van Zadelhoff et al.\ 2001), the systematic Keplerian
velocity field for an assumed stellar mass, the thermal line
width, the size of the disk, and the inclination of the
source. Qi (2000) estimates a disk inclination of $\sim 58\pm 10$
degrees, and an outer radius of the
CO disk of 435 AU for LkCa15.  In the following,
we assume a disk inclination of 60 degrees and a size (outer radius)
of 400 AU.
The dependence of line emission and profiles to disk radius
and inclination can be roughly estimated (Omodaka et al. 1992,
Beckwith \& Sargent 1993), e.g., for optically thick species the
peak intensities of molecular lines are proportional to the square of
the disk radius. Dependence of the integrated intensities on disk
inclination is also discussed in van Zadelhoff et al. (2001),
and is found to be less than a factor of two when the inclination is varied
from 0 to 60 degree.
%Line profiles with different disk radius and inclination
%can roughly be estimated referring to Omodaka et al. (1992) and
%Beckwith \& Sargent (1993), although these authors assumed constant molecular
%abundances relative to hydrogen. For example, peak intensities of molecular
%lines are in proportional to the square of the disk radius.
%Dependence of integrated intensities on
%disk inclination is also discussed in van Zadelhoff et al. (2001).

Fig.  \ref{fig:line400} compares simulated line profiles for CO
($J=6-5$, $3-2$, and $2-1$), HCO$^+$ ($4-3$), HCN ($4-3$), and CN
($3-2$) from the fiducial D'Alessio et al. model (thick solid
line) with lines from LkCa15 observed with single-dish telescopes
by van Zadelhoff et al. (2001) (solid line).
Line profiles based on the D'Alessio et al.\
model with other accretion rates -- $\dot{M}=10^{-9}$ (dot-dashed
line) and $\dot{M}=10^{-7}$ (dashed line) M$_{\odot}$ yr$^{-1}$ -- are
also shown.  In spite of the fact that the column densities of CO and
HCO$^+$ in our model are slightly smaller than those estimated by Qi
(2000), the calculated intensities of these species
are higher than observed in LkCa15 by a factor of $2-3$, which is caused by the
higher disk temperatures in our model. The vertical temperature gradient
lowers the optical depth of the disk, which further enhances emission
intensities (van Zadelhoff et al. 2001).
%It should also be noted that
%Qi (2000) estimated the CO column density from C$^{18}$O ($J=1-0$). We do not
%calculate the line here, because the line profile was not shown in Qi (2000).
As opposed to CO and HCO$^{+}$, the model
intensities of CN and HCN are much lower than those observed in
LkCa15. Those lines are optically thin in our model, and about 10 times
more HCN and at least 50 times more CN are needed to fit the observed profiles,
which is consistent with the comparison of column densities. The dotted
lines in the CN and HCN panels show model profiles in which the molecular
abundances are artificially enhanced.
We conclude that CN and HCN in LkCa15 are much more
abundant than in our model.

\subsection{Discussion}
What are the causes of the disagreement between our model results
and LkCa15, and the difference between DM Tau and LkCa15 ?
Let us first discuss the uncertainties in the estimate of the molecular
column densities and size of the emission region.
Although there are some differences in observed
intensities, the lines in LkCa15 are, in fact, not much stronger than those
in DM Tau, at most a factor of 2--3 after scaling the data to the same beam.
The main reason for the different column densities in DM Tau and LkCa15
then seems to be the adopted size of the emission region, because the
column densities are inversely proportional to the square of the size (radius)
of the emission region.
For CO, where interferometric observations have been performed both for
DM Tau and LkCa15, the DM Tau disk is found to be somewhat larger than
the LkCa15 disk --- DM Tau is about 800 AU in radius, while LkCa15 is 435 AU.
But whether this larger emission region in DM Tau holds for other
species is
not yet known without some degree of ambiguity because the
DM Tau data come mainly from single-dish observations,
and the size of the  region and average molecular abundances are
estimated from a model with constant abundances throughout the disk.
For LkCa15, in contrast, the sizes of the emission regions are directly
derived from interferometric observations and are at most a few arcseconds.
Assuming that the interferometer resolves the emission, these column densities
should therefore not be affected by uncertainties in the size of the emission
region. Although interferometry is more direct, and thus seems
more reliable for obtaining the size of the emission region,
it should be noted that the derived size depends on the signal to noise
ratio and dynamic range achieved in the observations.
In fact, the size of the emission region obtained via interferometry seems
to contain uncertainties; the size of the CO ($J=2-1$) emission region
around LkCa15 is estimated to be $\sim 600$ AU by Duvert et al. (2000),
which is larger than the 435 AU obtained by Qi (2000).

Since our 2D modeling procedure convolves the calculated molecular
emission with the actual beam of the observations, a larger disk size
cannot explain the discrepancy with the LkCa15 interferometer
observations, but it does affect the calculated single dish
intensities. Such an increase would not be sufficient to explain the
discrepancies for LkCa15, however.  For example, if we assume an outer disk
radius of 600 AU, the calculated HCN and CN line intensities from our
model disk (Fig. \ref{fig:line400}) are increased by
a factor of 2 at most, while the calculated CO and HCO$^+$ lines become
even stronger compared with the observed profiles.
There are a few possible
explanations for these discrepancies. First, our model might overestimate
the CO column density and underestimate the radical column density
because we do not consider dissociation of CO via stellar UV
radiation, as
mentioned above. Detailed consideration of stellar UV radiation,
including the CO
and H$_2$ dissociation, will be reported in a forthcoming paper.
Indeed, CN is enhanced by an order of magnitude depending on the treatment
of the radiation field, although HCN is not much changed.
The inclusion of X-rays might be another solution. X-rays cause ionization
and dissociation, which enhance chemical activity, and hence increase
the
transformation of CO to other organic molecules such as CN and HCN
(Aikawa \& Herbst 1999a, 2001). Different X-ray fluxes might also account
for the differing molecular column densities in DM Tau and LkCa15,
if they are intrinsic. X-rays also cause non-thermal desorption, which might
enhance the CN and HCN abundances in the gas phase (Najita, Bergin \& Ullom
2001). Finally, of the disks around T Tauri stars surveyed so far,
LkCa15 stands out as the disk with the strongest molecular lines
and richest chemistry in the interferometer and single-dish data
(see \S 4.1, Qi 2000, Thi et al. in preparation).

\section{Conclusion and Discussion}

       We have investigated the molecular distributions in protoplanetary
       disks by combining the ``new standard'' chemical model
       with the
       physical model of D'Alessio et al., which has a temperature
       gradient in the vertical direction.  The calculated molecular
       column densities are in reasonable agreement with those estimated
       from single-dish observations of the DM Tau disk without
       the assumptions in
       previous calculations of non-thermal desorption and/or an
       artificially low sticking probability. In the warmer intermediate layers
       of our disk models, there are large amounts of gaseous molecules
       owing to thermal desorption and efficient UV shielding caused by
       large gas densities at large heights from the midplane, a
       phenomenon known as flaring.  Gaseous molecules are abundant only
       in regions with certain physical conditions.  The volume of the
       layers with these conditions, and thus the column densities of
       gaseous molecules, are not proportional to the total (H$_2$) column
       density. Column densities of abundant
       molecules such as CN, HCN and HCO$^+$ do not vary by more than a
       factor of three during the period $t\sim 10^5-10^6$ yr. Sulfur-bearing
       molecules and H$_2$O show larger temporal variations.

       Comparison of our model results with those of Willacy
       \& Langer (2000), who adopted the C-G disk model with a Gaussian density
       distribution, indicates that gaseous molecular abundances are
       sensitive to the vertical structure of the disk model; in their
       model, molecules in the super-heated upper layer are destroyed by the
       harsh ultraviolet radiation from the star, while sufficient UV
       shielding is available in the warm upper layers of the D'Alessio et al.
       model.  Deuterated species are also included in our chemical model.
       The molecular D/H ratios we obtain are in reasonable agreement
       with those observed in protoplanetary disks.

Despite our agreement with observations of DM Tau, the molecular
column densities obtained in our models are smaller than those
observed around LkCa15, except for CO and HCO$^+$.  The estimated column
densities of all observed molecules around LkCa15 are higher than
those around DM Tau by roughly an order of magnitude. This
difference in derived molecular column densities in the two objects seems
to derive, at least partially, from different and/or uncertain sizes
of the emission regions. Comparison with other sources shows that some of
the difference is likely to be intrinsic, and other physical
parameters or processes, such as X-ray ionization and dissociation,
are needed to account for the high column densities in the LkCa15 disk.

In addition to the calculation of molecular abundances and column
densities, we have solved the equation of radiation transfer to obtain
line profiles from our model disks, which can be directly compared
with observations.  Such a comparison is a much more detailed test of
theory than is a comparison of column densities, since line
intensities depend not only on the molecular column densities but also
on the density and temperature of the molecular layer and the
variation of the abundance of the molecule with $R$ and $Z$.
The line intensities of HCN and CN obtained from the theoretical models
are lower than the observed intensities in LkCa15, as expected from the
comparison of column densities.

There are still several uncertainties in the vertical structure of
protoplanetary disks, which might affect our results.  Firstly, Chiang
\& Goldreich (1997) argue that the gas temperature could be lower than
the dust temperature in the upper layers; although an important heat
source of the gas is collisions with super-heated dust particles,
gas-dust collisions are not frequent enough to equilibrate the gas
temperature and grain temperature because of the low density.  With a
lower gas temperature, the disk would be less flared than in the model
of D'Alessio et al., which assumes equal temperatures of gas and dust.
Glassgold \& Najita (2001) have pointed out, however, that gases in the
upper layer can be heated by X-rays, which were not considered by
Chiang \& Goldreich (1997). Since Glassgold \& Najita (2001) have
listed the temperature only in the inner radius ($R=1$ AU), we made a
rough estimate of the disk surface temperature for the outer radii
$R\sim 100-300$ AU based on the work of Maloney et al.  (1996), which
suggests that the surface temperature of the X-ray irradiated disk is
indeed higher than the midplane (interior) temperature of the C-G
and Kyoto models. Moreover, the UV photons can heat the gas through the
photoelectric effect on grains and PAHs, as in models of photon-dominated
regions. Hence we
can at least expect disks to fall off more slowly in density with
increasing height than in a simple Gaussian distribution, although
detailed studies on the heating and cooling balance between gas and
dust are desirable in order to obtain an accurate vertical structure
for the disk.  Another uncertainty lies in the size and distribution
of dust particles.  D'Alessio et al.  (2001) and Chiang et al.  (2001)
have noted that their original models are geometrically too thick
compared with the observations of edge-on disks, which suggests dust
sedimentation and/or growth in the disk.  Because the molecular
abundances in our model depend on the efficiency of UV shielding by
``small'' (i.e.\ interstellar) dust grains (Aikawa \& Herbst 1999a),
we might have overestimated these abundances.  Although a more
detailed approach with dust sedimentation is beyond the scope of this
paper, we emphasize that molecular abundances can help to
resolve uncertainties in dust evolution and disk structure.

\acknowledgements
The authors are grateful to P.\ D'Alessio for providing
numerical tables of her models, to G.\ Blake, C.\ Qi and W.F.\ Thi for
results of their observations prior to publication, and to
M. Hogerheijde and F. van der Tak for use of their 2D Monte Carlo code.
Astrochemistry in Leiden is supported through a Spinoza grant from
the Netherlands Organization for Scientific Research (NWO).
Astrochemistry at Ohio State is supported through a grant from the
National Science Foundation. Numerical calculations were carried out at
the Astronomical Data Analysis Center of National Astronomical Observatory
in Japan.

\end{document}